\documentclass[twocolumn,preprintnumbers,superscriptaddress,amsmath,amssymb,nobalancelastpage,10pt,shortbibliography,noeprint]{revtex4-2}

\usepackage[utf8]{inputenc}
\usepackage{physics}
\usepackage{braket}
\usepackage{miller}
\usepackage{changes}
\usepackage{multirow}
\usepackage{float}
\usepackage{graphicx}
\usepackage{bm}
\usepackage{verbatim}
\usepackage{xfrac}
\usepackage{array}
\usepackage{siunitx}
\usepackage{ulem}
\usepackage{amssymb}
\usepackage{amsmath}
\usepackage{hhline}
\usepackage[hidelinks]{hyperref}
\usepackage{lineno}

\begin{document}
        \title{Exponentially enhanced sensing through non-reciprocal light propagation}

        \author{P.-É. Blanchard}
        \affiliation{Département de Physique, Université de Montréal, Montréal, Québec, Canada}
        
        \author{A. McDonald}
        \affiliation{Institut Quantique and Département de Physique, Université de Sherbrooke, Sherbrooke, Québec, Canada}
        
        \author{P. St-Jean}
        \email{philippe.st-jean@umontreal.ca}
        \affiliation{Département de Physique, Université de Montréal, Montréal, Québec, Canada}
        \affiliation{Institut Courtois, Université de Montréal, Montréal, Québec, Canada}

        \begin{abstract}
        Non-reciprocity is a key resource for pushing the performance of photonic devices beyond the fundamental limits imposed by Lorentz reciprocity. Notably, it was predicted that sensor architectures implementing a bosonic analogue of the Kitaev chain can yield an enhanced signal-to-noise ratio. However, experimental demonstrations of such sensors have so far been limited to small chains. Here we show how this model can be implemented in a scalable optical platform, enabling the detection of weak perturbations with a signal-to-noise ratio, linked to the quantum Fisher information extracted per input photon, that grows exponentially with system size. Our approach encodes two coupled Hatano-Nelson chains---realizing an effective bosonic Kitaev model---within the resonant modes of an electro-optic frequency comb. Non-reciprocal propagation in frequency space is achieved via simultaneous phase and amplitude modulation of the intracavity field. We demonstrate the sensing of a weak modulation tone coupling the two chains, observing an exponential enhancement of the signal-to-noise over three orders of magnitude as the system is scaled to over 70 frequency modes per chain. These results establish a new regime of non-Hermitian sensing, highlighting how non-reciprocal dynamics can be harnessed to achieve exponentially enhanced measurement precision in photonic systems.
        \end{abstract}
        \maketitle

Photonic devices rely on the controlled manipulation of electromagnetic fields through a combination of passive processes, such as engineered interference and dissipation, and active processes, including gain or parametric interactions~\cite{lipson_guiding_2005, marpaung_integrated_2019, cui_roadmap_2024, obrien_photonic_2009, slussarenko_photonic_2019, wang_integrated_2020}. Fundamental constraints on these processes are imposed by Lorentz reciprocity, which restricts the directional flow of energy and information in linear, time-reversal-symmetric systems. In recent years, substantial efforts have therefore focused on realizing nonreciprocal photonic platforms -- both classical and quantum -- with the aim of surpassing these limits~\cite{sounas_non-reciprocal_2017, barzanjeh_nonreciprocity_2025}. Beyond their established role in optical isolation, nonreciprocal systems have been shown to enable qualitatively new operating regimes in photonic devices, notably in the contexts of topological lasers~\cite{bahari_nonreciprocal_2017, bandres_topological_2018, klembt_exciton-polariton_2018}, nonreciprocal thermal radiation~\cite{yang_nonreciprocal_2024}, quantum-limited amplification~\cite{kamal_noiseless_2011} and chiral quantum optics~\cite{lodahl_chiral_2017, owens_chiral_2022, suarez-forero_chiral_2025}.

It has been theoretically proposed that nonreciprocal light propagation can similarly provide a substantial advantage in optical sensing~\cite{metelmann_quantum-limited_2014, lau_fundamental_2018, bao_fundamental_2021}. One key architecture to achieve this advantage is based on a bosonic analogue of the Kitaev model~\cite{mcdonald_phase-dependent_2018, mcdonald_exponentially-enhanced_2020}, in which nonreciprocal dynamics cause signal and noise to scale differently with system size, enabling an exponential enhancement of the signal-to-noise ratio (SNR), linked to the quantum Fisher information extracted per input photon. However, realizing this model requires specific two-photon processes -- reminiscent of Cooper-pair formation in the fermionic Kitaev model -- that are intrinsically challenging to implement and scale. As a result, experimental realizations have so far been limited to small lattices comprising only 3-5 sites~\cite{busnaina_quantum_2024, slim_optomechanical_2024}. Moreover, none of these earlier works studied the scaling of noise, a critical prerequisite for assessing a sensor’s genuine performance.

In this work, we present and demonstrate a technique to achieve this exponential scaling of the SNR. Our platform consists of two coupled Hatano-Nelson (HN) lattices~\cite{hatano_localization_1996} -- the canonical nonreciprocal lattice Hamiltonian -- encoded in an electro-optics frequency comb, where each frequency mode acts as a lattice site~\cite{yuan_synthetic_2021}. Propagation along each HN lattice is equivalent to the chiral propagation of the field's quadratures in a bosonic Kitaev chain (BKC). Nonreciprocal propagation in frequency space is achieved by simultaneously modulating the phase and amplitude of the circulating field in the optical fiber cavity~\cite{wang_generating_2021}. This frequency-encoding scheme allowed us to demonstrate a distinct amplification of the signal and noise, leading to an exponential scaling of the SNR in combs formed from up to 70 lattice sites. This exponential advantage opens up pathways for engineering non-hermitian sensors with enhanced performances, in particular for applications highly sensitive to environmental noise.\\

\begin{figure}
	\centering
	\includegraphics[width=0.9\linewidth]{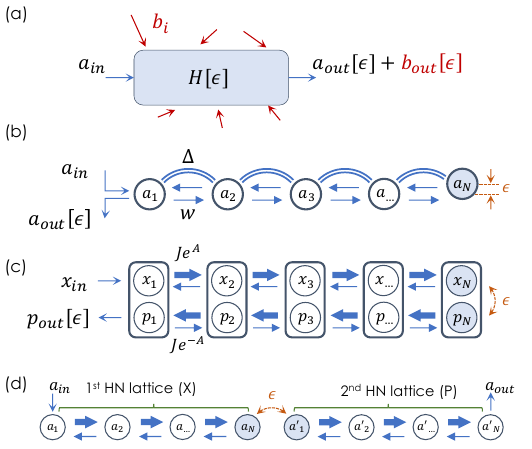}
	\caption{\textbf{Optical sensing with a bosonic Kitaev chain.} (a) Schematic depiction of a basic optical sensors transforming an input field $a_{in}$ into an output field $a_{out}$ whose properties depend on an external perturbation $\epsilon$. Mitigating the sensitivity to external noise ($b_i$, red arrows), which leads to a noisy output field ($B_{out}$), is the key challenge of optical sensors. (b) Depiction of a BKC where every lattice site ($a_i$) is coupled to its nearest-neighbor through one-($w$) and two-($\Delta$) photon processes.  (c) The BKC is equivalently described by the chiral propagation of light's quadrature ($x$ and $p$) with hopping amplitudes $Je^{\pm A}$. Lattices sites are depicted as rectangles. (d) This model is similar to two coupled HN lattices where each simulates a given quadrature. The system is driven on the $1^{st}$ lattice of the $X$-chain and readout by probing the last site of the $P$-chain.}
	\label{fig:figure1}
\end{figure}

\noindent{\em Nonreciprocal sensing with a BKC --} At the basic level, an optical sensors consists of a system, described by a Hamiltonian $H$, that is sensitive to a specific family of external perturbations $\epsilon$ (see Fig.~\ref{fig:figure1} (a)). This system can take an input optical field $a_{in}$ and will then output a field $a_{out}$ whose properties depend on both $\epsilon$ and $a_{in}$, allowing to infer the nature and magnitude of $\epsilon$. The quantum Fisher information (QFI) quantifies the maximal information on the system that can be extracted from the probe field $a_{out}$~\cite{degen_quantum_2017,giovannetti_advances_2011}. Optical sensors are, inherently, open quantum systems~\cite{pirandola_advances_2018, frascella_overcoming_2021}; they are thus sensitive to environmental electromagnetic noise ($b_{i}$) entering the device, leading to a deterioration of the SNR in the output field and therefore of the QFI that can be extracted per photon injected in the system.

\begin{figure*}
	\centering
	\includegraphics[width=1\linewidth]{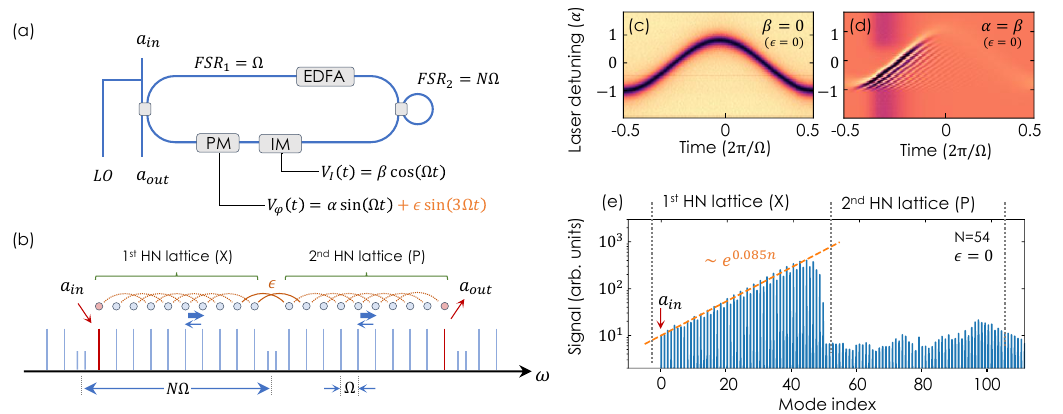}
	\caption{\textbf{Non-Hermitian skin effect in frequency space.} (a) Schematic representation of our photonic platform formed from a main optical fiber loop with FSR $\Omega$ coupled to a smaller loop with a FSR $N\Omega$. An erbium-doped optical amplifier (EDFA) is added to the cavity to mitigate losses. The circulating field is modulated in phase (PM) and intensity (IM) at frequency $\Omega$ using electro-optic modulators driven by voltages $V_{\varphi}(t)$ and $V_{I}(t)$, respectively, to induce nonreciprocal light propagation in frequency space. An optical amplifier (EDFA) is added to the loop to mitigate losses. The input signal is split into two paths to provide a local oscillator (LO) for the heterodyne measurement. (b) The resulting frequency spectrum consists of a succession of $N$ resonant modes separated by $\Omega$; we only consider two of these arrays to form HN lattices (emulating the $X$ and $P$ quadratures of the BKC). The driven and probed modes are depicted in red, each belonging to a distinct HN chain coupled by a perturbation $\epsilon$. The perturbation consists of a reciprocal $3^{rd}$ nearest-neighbor coupling term; the solid (dashed) lines depict couplings between (within) each chain. (c)-(d) Time-resolved transmission measurements allows measuring the effective band structure for the reciprocal ($\beta=0$) and nonreciprocal ($\alpha=\beta$) propagation. (e) Heterodyne spectrum (for the nonreciprocal case with $\epsilon=0$) allowing to probe the distribution of the field in frequency space; each HN lattice is separated by gray dotted lines. A clear exponential accumulation associated to the NHSE (oranged dashed lines) is observed in the driven lattice; there is no significant emission in the $2^{nd}$ lattice.}
	\label{fig:figure2}
\end{figure*}

One key approach for overcoming this fundamental limitation is to engineer systems whose response ($a_{\rm out}$) to the input driving field scales differently from their response ($b_{\rm out}$) to spurious noise sources, thereby enhancing the output SNR ($|a_{\rm out}|/|b_{\rm out}|$). This is in contrast to several earlier designs inspired by non-Hermitian topology, which primarily enhance an observable response but do not change how signal and noise scale relative to one another. Exceptional-point sensors, for example, can exhibit greatly enhanced eigenvalue sensitivity to perturbations via coalescing eigenmodes and engineered gain~\cite{miri_exceptional_2019, chen_exceptional_2017, hodaei_enhanced_2017, wiersig_review_2020}, but both signal and noise are typically amplified similarly, so there is no net gain in terms of the extracted QFI per photon~\cite{lau_fundamental_2018, langbein_no_2018}. As well, implementations based on propagating the field in single nonreciprocal lattices~\cite{budich_non-hermitian_2020, parto_enhanced_2025} where the non-Hermitian skin effect (NHSE) allows generating large spectral or intensity shifts at the boundaries can lead to exponentially-enhanced sensitivities. However, when all ports and noise channels are treated consistently, these single-chain architectures also do not exhibit a distinct scaling of signal and noise, and the QFI per input photon does not grow with system size~\cite{lau_fundamental_2018, mcdonald_exponentially-enhanced_2020, bao_fundamental_2021}.

As proposed in Ref.~\cite{mcdonald_exponentially-enhanced_2020}, architectures inspired by a bosonic analogue of the Kitaev chain can provide an exponential enhancement of the SNR with system size. As depicted in Fig.~\ref{fig:figure1} (b), a BKC is formed from a one-dimensional array of N resonators coupled by single-particle hopping ($w$) and two-mode squeezing ($\Delta$) processes. Its physics is captured by the Hamiltonian:
\begin{equation}
	H_{BKC} = \sum_j \left(i w a^{\dagger}_{j+1}a_{j} + i \Delta a^{\dagger}_{j+1}a^{\dagger}_{j} + \rm{h.c.}\right) .
\end{equation}
where $a_{i}$ and $a_{i}^{\dagger}$ are the canonical annihilation and creation operators on the $i^{th}$ site, and h.c. is the hermitian conjugate. The perturbation to be sensed in this implementation consists of a small on-site detuning of the last resonator $\epsilon a^{\dagger}_{N}a_{N}$. The system is probed with a homodyne measurement through an input-output waveguide coupled to the first site which, in the limit of a strong driving field, is optimal for extracting the QFI in a linear, Gaussian system~\cite{lau_fundamental_2018, pinel_quantum_2013}.

To better understand how this Hamiltonian allows sensing the presence of $\epsilon$ with a SNR that grows exponentially with $N$, it is insightful to express it with the quadrature operators $x_{i}=(a_{i}^{\dagger}+a_{i})/\sqrt{2}$ and $p_{i}= i(a_{i}^{\dagger}-a_{i})/\sqrt{2}$:
\begin{equation}
	H_{BKC} = \sum_i Je^{A} p_{i+1}x_{i} - Je^{-A}x_{i+1}p_{i}
\end{equation}
where $Je^{\pm A}=w\pm\Delta$.

This latter Hamiltonian preserves the BKC's $\mathbb{Z}_{2}$ symmetry and describes the counter-propagation of the two quadratures (Fig.~\ref{fig:figure1} (c)) with the $X$ ($P$) hopping preferentially to the right (left). In this equivalent framework, $\epsilon$ breaks the system's $\mathbb{Z}_{2}$ symmetry by mixing the two quadratures akin to a magnetic impurity in the quantum spin Hall effect that mixes counter-propagating electrons with opposite spins~\cite{maciejko_quantum_2011}. As such, driving the lattice with a coherent field in the $X$ quadrature leads to an exponential enhancement of its amplitude as it propagates to the other edge of the chain through the so-called non-hermitian skin effect (NHSE)~\cite{okuma_topological_2020}; there, part of the signal, upon the effect of $\epsilon$, is transferred to the $P$ quadrature and is thus re-amplified as it propagates back to the input-output waveguide. In contrast, at lowest order in $\epsilon$, the noise being homogeneous in quadrature is not similarly amplified. This leads to a QFI per photon -- which here coincides with the SNR in the limit $\epsilon \to 0$ -- that scales exponentially with $A$ (the nonreciprocity) and $N$ (the system size)~\cite{lau_fundamental_2018}. The exponential scaling of this fundamental metric demonstrates the genuine sensing advantage of this model.


To implement this scheme, we consider an equivalent model formed from two copies of the HN model -- i.e. lattices with nonreciprocal nearest-neighbor hopping terms - where each HN lattice captures the evolution of a given quadrature in a BKC (see Fig.~\ref{fig:figure1} (d)). The perturbation $\epsilon$ is implemented by simply connecting the last site of the first lattice to the first site of the second lattice. This implementation is much easier to scale as it does not require two-mode squeezing processes at each site. The noise considered in this work solely comes from the environment, i.e. thermal noise populating each lattice site. It does not take into account the inherent shot noise of the input drive that necessarily leads to exponentially-enhanced fluctuations in the output field. Mitigating this latter contribution would require driving the system with squeezed light rather than a coherent field. This however raises the question of how significant amounts of squeezing can be maintained in a non-hermitian environment with losses~\cite{sloan_noise_2025}.


It is important to point out that although the SNR can be improved by either increasing the magnitude of the nonreciprocity ($A$) or the system size ($N$), the latter is more advantageous in our double-HN implementation. Indeed, nonreciprocal propagation of the field's amplitude is a non-hermitian feature that requires coupling to the environment~\cite{metelmann_nonreciprocal_2015, clerk_introduction_2022}; this necessarily leads to enhanced noise in the device through the fluctuation-dissipation theorem~\cite{clerk_introduction_2010}. In contrast, scaling $N$ allows working  with very small nonreciprocal factors, hence relatively weak couplings to the environment. This is the approach that we favor in this work.\\

\noindent{\em Coupled HN lattices in frequency space --} Our implementation of a double HN lattice architecture, as depicted in Fig.~\ref{fig:figure1} (d), uses an approach based on the concept of electro-optics frequency combs~\cite{parriaux_electro-optic_2020}. The platform consists of a main optical fiber loop cavity (Fig.~\ref{fig:figure2} (a)) where each frequency eigenmode embodies a distinct lattice site. Hence the mode periodicity in frequency given by the cavity's free spectral range (FSR) mimics the spatial periodicity of a conventional solid-state lattice. An optical amplifier is embedded in the cavity to mitigate losses and extend the lifetime. The cavities' FSR is varied between $1-10~\mathrm{MHz}$ to study different lattice sizes.

\begin{figure*}
	\centering
	\includegraphics[width=1\linewidth]{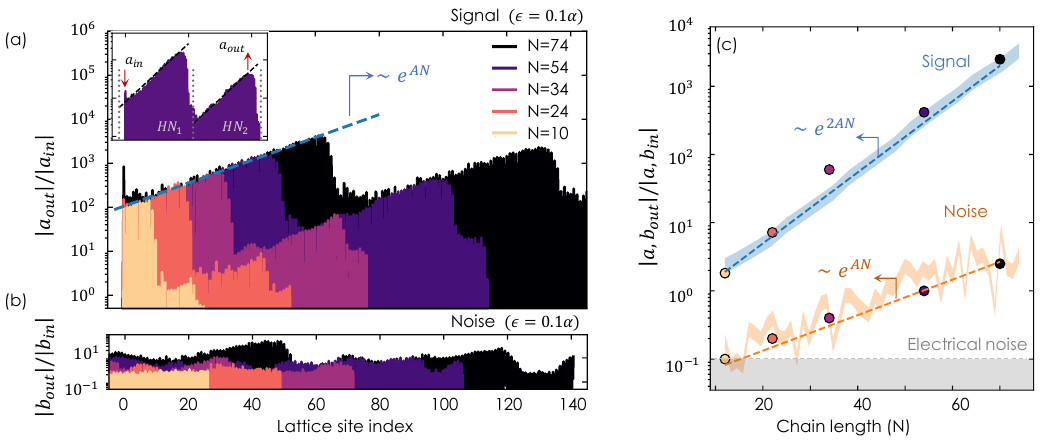}
	\caption{\textbf{Exponential enhancement of the SNR.} (a) Heterodyne spectra, normalized by the input field amplitude $|a_{in}|^{2}$, for different lengths of the main optical fiber loop leading to HN lattice sizes of $N=10,24,34,54,74$. Each measurement is taken with a coupling perturbation $\epsilon=0.1\alpha$. The inset is a close-up to the case $N=54$ that shows a clear amplification $e^{AN}$ (dark dashed lines, blue dashed line in the main part) in both HN chains separated by the gray dotted lines. The frequency of the driving field and probed modes are indicated by red arrows. (b) Similar measurement when the laser is not resonant with the main loop to obtain noise spectra. Panel (c) summarizes the scaling of the readout signal and noise exhibiting a clear distinction between the scaling of the signal ($e^{2AN}$, blue dashed line) and noise ($e^{AN}$, orange dashed line). The steady-state solution of the Langevin equation is shown by shaded areas whose width indicates the experimental uncertainty associated to the measured RMS of the gain and inherent fluctuations of the input field included in the simulations. The gray area depicts the photodiode's electrical noise limit.}
	\label{fig:figure3}
\end{figure*}

In order to obtain finite-sized lattices, we couple this main cavity to a secondary, smaller optical fiber loop with a FSR of $N\Omega$. As a result, one out of every $N$ modes of the main cavity couples with a mode of the small loop, leading to frequency-split supermodes~\cite{dutt_creating_2022}, which effectively breaks the main frequency combs. The system's eigenspectrum thus exhibits a succession of finite-sized, $N$-teeth frequency combs split by $2\Omega$ (Fig.~\ref{fig:figure2} (b)). Two of these combs will host the HN lattices required for implementing the BKC analogue depicted in Fig.~\ref{fig:figure1} (d).

To implement the HN lattices, nonreciprocal coupling between neighboring frequency modes is achieved using a combination of electro-optical phase (PM) and intensity (IM) modulators driven at a frequency $\Omega$ and with a $\pi/2$ phase difference~\cite{wang_generating_2021}. Controlling the PM and IM drive amplitude independently ($\alpha$ and $\beta$ respectively) allows tuning precisely the nonreciprocity $A$ of light propagation in frequency space. Although there is theoretically no limits on the nonreciprocity achievable, we don’t work in regimes of complete directionality because the amplification of the field would lead to nonlinear effects in the fiber. The perturbation that couples neighboring HN lattices consists of a small Fourier component added to the PM at frequency $3\Omega$ and with amplitude $\epsilon$ (Fig.~\ref{fig:figure2} (b)). In our proof-of-concept experiment, this perturbation fits naturally within our setup but, in a realistic sensing context, the same mechanism could be functionalized to optically detect weak microwave tones via electro-optic processes. We did not use a perturbation at $2\Omega$ because the first and last sites of each lattice slightly couple with the small loop leading to a decrease of their lifetime. It is important to point out that the perturbation also couples each site to its third-nearest neighbor (dotted lines). This departure from the canonical model presented in Fig.~\ref{fig:figure1} (d) does not impact qualitatively the overall performance of the device (see Supplementary Section IV for details).

The band structure of each lattice can be obtained by performing time-resolved transmission measurement by scanning a continuous-wave laser through any cavity mode of a lattice, using an under-coupled transmission line and a high-bandwidth photodiode~\cite{dutt_experimental_2019, chenier_quantized_2026}. If we only modulate the main loop in phase ($\beta=0$, Fig.~\ref{fig:figure2} (c)), we obtain a completely symmetric band structure indicating that the couplings are fully reciprocal. In contrast, by modulating both the phase and intensity ($\alpha=\beta$) of the circulating field, we obtain a band structure whose intensity is highly asymmetric, with a dominant occupation of modes with a rightward propagating group velocity (as seen in Fig.~\ref{fig:figure2} (d)). This asymmetry is a clear signature of the nonreciprocal couplings between lattice sites~\cite{wang_generating_2021}. The observed ripples on the right-hand side of the rising part of the dispersion indicate an occupation in multiple neighboring chains. This inter-chain leakage, caused by the second-order sidebands of the $\Omega$ phase modulation, is apparent even for $\epsilon=0$ as this specific measurement is realized with a strong amplification.

To provide an intuitive picture of how simultaneous phase and intensity modulations create non-reciprocal propagation, we can look at their individual effects. First, the phase modulation periodically changes the resonating frequencies of the main loop, leading to band dispersions where time acts as the crystal momentum (Fig.~\ref{fig:figure2} (c)). Second, the out-of-phase intensity modulation increases the circulating field when the dispersion has a positive slope -- meaning the field decreases in frequency -- and reduces it when the slope is negative (Fig.~\ref{fig:figure2} (d)).

We then realize a frequency-space spectroscopy of the output field by beating the signal radiating from the main loop with a frequency-shifted copy of the driving laser~\cite{dutt_single_2020}. This heterodyne measurement allows extracting the amplitude of each frequency component of the field. Working in a regime where $\epsilon=0$ and $\beta/\alpha\sim0.09$, leading to a ratio of $\sim0.83$ between leftward and rightward hopping terms ($A=0.085$), we obtain the frequency profile depicted in Fig.~\ref{fig:figure2} (e). The dotted gray lines indicate the edges of each lattice. We clearly observe an exponential localization of the field toward the higher frequencies of the pumped lattice, as expected from the NHSE~\cite{okuma_topological_2020, weidemann_topological_2020}. Here, compared to Fig.~\ref{fig:figure2} (d), we have lowered the amplification to ensure that only one chain is significantly occupied.\\

\noindent{\em Exponential enhancement of the SNR --} We now implement a perturbation coupling the two lattices by adding a small Fourier component at frequency $3\Omega$ with an amplitude $\epsilon =0.1\alpha$ and measure the heterodyne spectrum as previously described. We perform this measurement for different sizes of the main loop, leading to lattice sizes of $N=10,~24,~34,~54,~74$ frequency modes (Fig.~\ref{fig:figure3} (a)). We here use a reduced nonreciprocal factor $A=0.06$ to mitigate nonlinear effects in the cavity. For each configuration, we now observe a clear NHSE in both lattices. The output signal ($a_{out}$) is measured as the amplitude of the Fourier component at the edge of the second lattice (see inset in Fig.~\ref{fig:figure1} (a)). The result, normalized by the driving field amplitude $|a_{in}|$, is plotted in Fig.~\ref{fig:figure3} (c) as a function of $N$ -- we can change $N$ by changing the FSR of the main loop (see Methods and Supplementary Information, section I for details). We clearly see a $e^{2AN}$ exponential scaling of the output field's intensity, in agreement with the fact that the output signal is measured $2N$ sites away from the injection frequency.

We then assess the system's response to environmental noise to extract the SNR. In comparison to lattices encoded in real space (as in Ref.~\cite{mcdonald_exponentially-enhanced_2020} and Fig.~\ref{fig:figure1} (b)-(d)) where only the first lattice site is coupled to the input-output waveguide, in our frequency-encoding scheme (as in Ref.~\cite{slim_optomechanical_2024}) all modes are identically coupled to the environment. Therefore, noise is injected identically in all frequency modes, and to extract its propagation in the system, we measure the heterodyne signal without the input laser. In such a regime, the photonic field in the loop only comes from the amplified spontaneous emission (ASE) of the optical amplifier embedded in the loop. It thus produces an incoherent white noise that drives with equal amplitude every cavity mode. The advantage of using ASE as our source of noise is twofold : 1- it accurately describes the noise generated by the amplifier in our specific implementation, 2- it is broad enough spectrally to provide an accurate depiction of environmental white noise. 

The heterodyne signal for each chain length is plotted in Fig.~\ref{fig:figure3} (b). The scaling of noise as a function of $N$, extracted as the maximal value of the heterodyne signal throughout the entire spectra, is plotted in Fig.~\ref{fig:figure3} (c). Contrary to the scaling of the signal, we now observe a $e^{AN}$ scaling, and thus a $e^{AN}$ enhancement of the SNR. We clearly see that the two HN chains used in our implementation (which mimic the $\mathbb{Z}_{2}$ symmetry of the BKC model) are crucial for reaching this exponential scaling of the SNR and therefore of the QFI extracted per probe photon. Note that this does not violate the Heisenberg limit as the scaling is exponential in system size $N$ and not in the number of input photons.

\begin{figure}
	\centering
	\includegraphics[width=0.9\linewidth]{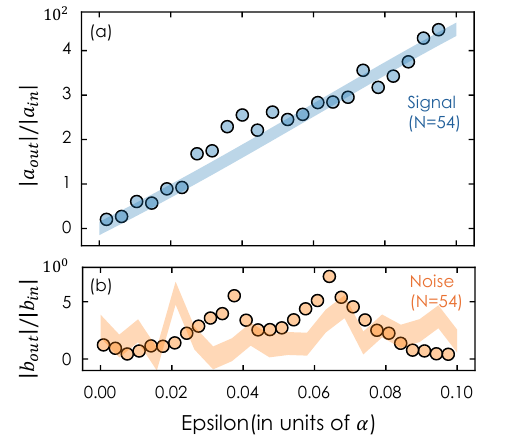}
	\caption{\textbf{Scaling of the SNR as a function of $\epsilon$.} Output signal (a, blue curve) and noise (b, orange curve) as a function of the perturbation strength $\epsilon$ for a given chain length $N=54$. The vertical axis of Panel (b) is two orders of magnitude smaller than that in (a). The shaded lines indicate the steady-state solutions of the Langevin equation with their width indicating the uncertainty calculated similarly as in Fig.~\ref{fig:figure3} (b).}
	\label{fig:figure4}
\end{figure}

This SNR scaling is distinct from the $e^{2AN}$ scaling reported in Ref.~\cite{mcdonald_exponentially-enhanced_2020}. This is explained by the fact that noise enters identically through every lattice site rather than preferentially through the first one. In the latter case~\cite{mcdonald_exponentially-enhanced_2020}, the output noise is independent of $N$ (in the limit $\epsilon\rightarrow 0$) and is simply given by the thermal occupation of the input-output waveguide. This scaling of noise in our system does not come from a redistribution across different channels as is often the case in non-hermitian sensing~\cite{lau_fundamental_2018,bao_fundamental_2021}. Rather, it is linked to the homogeneous spectral profile of noise that fundamentally differ from the localized probe field, leading to a distinct scaling. An intuitive picture to understand this is linked to the translational invariance of our system and of the noise profile in frequency space: noise entering each chain from the left (by hopping from its leftward neighbor) is exactly compensated by noise exiting to the right (see Supplementary Section II and III for details and analysis).

We further demonstrate this scaling of noise by performing a time evolution of the Langevin equation:
\begin{equation}
	\partial_{t} a_{i} = i[H(t), a_{i}]-\frac{\gamma}{2}a_{i}-i\sqrt{\kappa}a_{in}^{(i)}(t)
\end{equation}
with $a_{i}$ the field amplitude in each frequency mode, $H(t)$ the Hamiltonian describing light propagation in the double HN lattice, $\kappa$ the input-output coupling strength and $a_{in}^{(i)}$ injected field in each frequency mode. The amplifier is operated well below the lasing threshold, so that the effective loss rate $\gamma$ is given by the passive cavity losses minus its gain. In this regime, the steady-state propagation of signal and ASE noise can be described by the same Langevin equation with different inputs (coherent vs stochastic). To model the thermal noise injected in our system, we consider a $\delta$-correlated input field with
\begin{equation}
	\left\langle a^{\dagger(i)}_{in}(t)a_{in}^{(j)}(t')\right\rangle = \bar{n}_{th}\delta_{ij}\delta(t-t'),
\end{equation}
where $\bar{n}_{th}$ is the average number of photons spontaneously emitted by the amplifier. 

The result, when looking at the noise maximal intensity in the readout chain when the system reaches its steady-state, is plotted in Fig.~\ref{fig:figure3} (c) (shaded orange area where the width accounts for the experimental uncertainty on $A$ and $\epsilon$). We see a perfect agreement with our experimental data, both for noise and signal (shaded blue area, obtained with a localized coherent pump), hence confirming the $e^{AN}$ scaling of the SNR, and thus the QFI per input photon. Interestingly, this distinct amplification of the signal and noise also leads to an exponential enhancement of the sensitivity through a scaling of the noise-equivalent power $\mathrm{NEP}\propto e^{-AN}$. 

Finally, we turn to the evolution of the output signal and noise as a function of perturbation strength $\epsilon$, for a constant chain length $N=54$. The results are plotted in Fig.~\ref{fig:figure4}. We clearly see a linear increase of the signal's amplitude (a) as expected at such first-order in $\epsilon$, but the noise amplitude (b) in contrast remains small and constant. Both of these scalings match very well the Langevin equation steady states (shaded curves in Fig.~\ref{fig:figure4}), confirming these experimental observations. This is a clear demonstration that, in our specific implementation, signal and noise maintain a distinct scaling over a large range of perturbation strengths $\epsilon$. Importantly, it also demonstrates that the SNR scaling remains the same in the limit $\epsilon\rightarrow0$.\\ 

\noindent{\em Conclusion -- } We have demonstrated a nonreciprocal sensing architecture inspired by a bosonic analogue of the Kitaev model, in which signal and noise scale differently with system size, leading to an exponential enhancement of the SNR, linked to the QFI per input photon. While the perturbations investigated here are artificially engineered, our platform can be naturally extended to realistic sensing scenarios. In particular, optical fiber cavities -- unlike nano- and micro-resonators -- are well suited for remote optical sensing applications~\cite{passaro_recent_2012, butt_review_2023}. Moreover, since our approach relies on optically detecting weak microwave signals, it could provide a route to efficient, high-fidelity optical readout of superconducting qubits~\cite{mirhosseini_superconducting_2020, delaney_superconducting-qubit_2022, van_thiel_optical_2025, arnold_all-optical_2025, warner_coherent_2025} -- a task known to be highly sensitive to environmental noise~\cite{lauk_perspectives_2020, han_microwave-optical_2021}.

\vspace{20pt}

\noindent\textbf{Acknowledgements.} We acknowledge insightful discussions with I. Carusotto, G. Villa, O. Zilberberg, E. Verhagen, Clément Fortin and Tami Pereg-Barnea.

\vspace{20pt}

\noindent\textbf{Funding} PEB and PSJ acknowledge financial support from Québec's Fonds de Recherche--Nature et Technologies (FRQNT), Canada's Natural Sciences and Engineering Research Council (NSERC), the Alliance Quantum Program grant funded by NSERC for the project entitled “A new generation of hardware efficient superconducting qubits” and Québec's Minstère de l'Économie, de l'Innovation et de l'Énergie. AM
acknowledges funding from NSERC and the Canada First Research Excellence Fund.

\vspace{20pt}

\noindent\textbf{Author Contributions} PEG conducted the experiment, analyzed the data and run all the simulations. AM provided the theoretical framework of the BKC sensing scheme. PSJ designed the experiment, led the project and wrote the manuscript, with inputs from all authors.

\vspace{20pt}

\noindent\textbf{Competing Interests} The authors declare no competing interests.

\vspace{20pt}

\noindent\textbf{Data availability} All data used within this Article are available via figshare at \url{https://doi.org/10.6084/m9.figshare.33935710}.~\cite{DA}

\vspace{20pt}

\noindent\textbf{Code availability} Simulation codes are available via figshare at \url{https://doi.org/10.6084/m9.figshare.33935710}.~\cite{DA}

\clearpage
\noindent\textbf{Methods.}\vspace{5pt} \\
\textit{Optical fiber loop setup.} The entire platform is formed from polarization maintaining fibered components with the fast axis of the fibers oriented along the main axis of the lithium niobate crystal in the electro-optical phase and intensity modulators. The CW and CCW circulating modes of the main and auxiliary fiber-loops, respectively, are coupled with a 25:75 fiber coupler (FC). A circulator is further added to the main loop to ensure that we only consider a single propagation direction. To study HN chains of different sizes, we modify the main loop's FSR using various delay lines with length ranging from $0$ to $\SI{100}{\meter}$. For all system sizes, the finesse of the cavity remains identical, ensuring that the average number of passage in the amplifier remains identical. The spontaneous emission of the amplifier is restricted to a bandwidth of $\SI{100}{\giga\hertz}$ around the laser frequency, using a dense wavelength division multiplexing filter (DWDM). This allows justifying the assumption of a delta-correlated white noise as the coherence time of the amplifier ($<\SI{10}{\pico\second}$) is much shorter than any other relevant timescale in our experiment.

The nearest-neighbor couplings are induced by driving the phase (PM) and intensity (IM) modulators with, respectively, the following voltages:
\begin{equation}
\label{eq:coupling_nn}
\begin{aligned}
    &V_{\varphi}(t) = \alpha~\mathrm{sin}(\Omega t)\\
    &V_{I}(t)    = \beta~ \mathrm{cos}(\Omega t)
\end{aligned}
\end{equation}
with $\Omega$ the main loop's FSR. An alternative, intuitive picture to understand how this simultaneous intensity and phase modulation gives rise to a nonreciprocal propagation in frequency space comes from the fact that each process creates spectrally symmetric sidebands that are either in phase (IM) or out of phase (PM). Therefore, their simultaneous effect is to produce a constructive or destructive interference at lower or higher frequency. Here, we do not use identical modulation strengths, hence incomplete interference, to control the level of non-hermitian amplification in each chain. Moreover, we work in a regime where the bands' linewidth is limited by the cavity lifetime, indicating that RF noise is negligible.

The perturbation to be probed consists of and additional tone added to the PM:
\begin{equation}
    V_{\epsilon}(t) = \epsilon~\mathrm{sin}(3\Omega t).
\end{equation}

All modulation amplitudes ($\alpha$, $\beta$ and $\epsilon$) have units of Volts, leading to a modulation of the loop dictated by their respective electro-optical coefficients. They are calibrated by measuring the band structure with only one tone at a time. By doing so, the amplitude of the band structures measured indicates the strengths of each coupling tone. This procedure needs to be reiterated each time the FSR of the main loop is changed, because the electro-optics coefficient of the modulators change with the driving frequency.\\

\noindent\textit{Transmission measurements.}
The measurement of band structures in Fig.~\ref{fig:figure2} is performed by scanning a narrow-linewidth ($\Delta\omega<\SI{100}{\hertz}$) continuous-wave laser and probing the transmitted signal with a high-bandwidth photodiode. The entire frequency scan must be fast enough to ensure that the effect of thermal fluctuations in the loop remain small enough, but slow enough to allow the system to reach its steady-state : we typically scan the laser frequency over $\SI{4}{\mega\hertz}$ per ms. The frequency scan was necessary because the loop was not actively stabilized ; this could be circumvented using Pound-Drever-Hall locking .

The transmitted intensity measured on the photodiode ($I_{PD}(t)$ exhibits periodic Lorentzian dips, with period $T=2\pi/\Omega$, whenever the laser is resonant with the lower/upper band dispersions $E(k)$:
\begin{equation}
    \frac{I_{PD}(t)}{|a_{in}|^{2}}\simeq \left. \left[ 1-\frac{\kappa}{2}\mathrm{Re} \left\{ \frac{1}{\gamma/2-i(\Delta_{L}-E(k))} \right\} \right] \right|_{k=t~\mathrm{mod}~T} 
\end{equation}
where$\kappa$ is the input-output coupling strength of the cavity and $\Delta_{L}$ is the laser detuning with respect to the closest cavity mode. Here, as in other synthetic frequency dimension schemes (see Ref.~\cite{dutt_experimental_2019, pellerin_wave-function_2024, chenier_quantized_2026}), the crystal momentum $k$ has units of time, because our lattices are periodic in the frequency dimension. At each time $t$, the fields probe the eigen state of the Hamiltonian with crystal momentum $k=t~\mathrm{mod}~T$ within an effective Brillouin zone of size $T$, leading to a Lorentzian intensity dip whenever the laser frequency matches that of the corresponding eigenstate $E(k)$.\\

\noindent\textit{Heterodyne measurement.} In order to remove the large signal coming from the transmitted laser, the heterodyne measurements are performed by collecting light radiating from a distinct output port (not shown in Fig.~\ref{fig:figure1} than the input port used for injecting light. This signal is then mixed with the frequency-shifted ($\SI{200}{\mega\hertz}$) laser field.

\clearpage
\onecolumngrid

\setcounter{equation}{0}
\setcounter{figure}{0}
\setcounter{table}{0}
\setcounter{section}{0}
\renewcommand{\theequation}{S.\arabic{equation}}
\renewcommand{\thefigure}{S\arabic{figure}}
\renewcommand{\thetable}{S\arabic{table}}


\begin{center}
    \large{\textbf{\textsc{Supplementary Material: Exponentially enhanced sensing \\through nonreciprocal light propagation}}}
\end{center}

\section{Experimental method}

\subsection{Experimental setup}

\begin{figure}[H]
\centering
\includegraphics[totalheight=6cm]{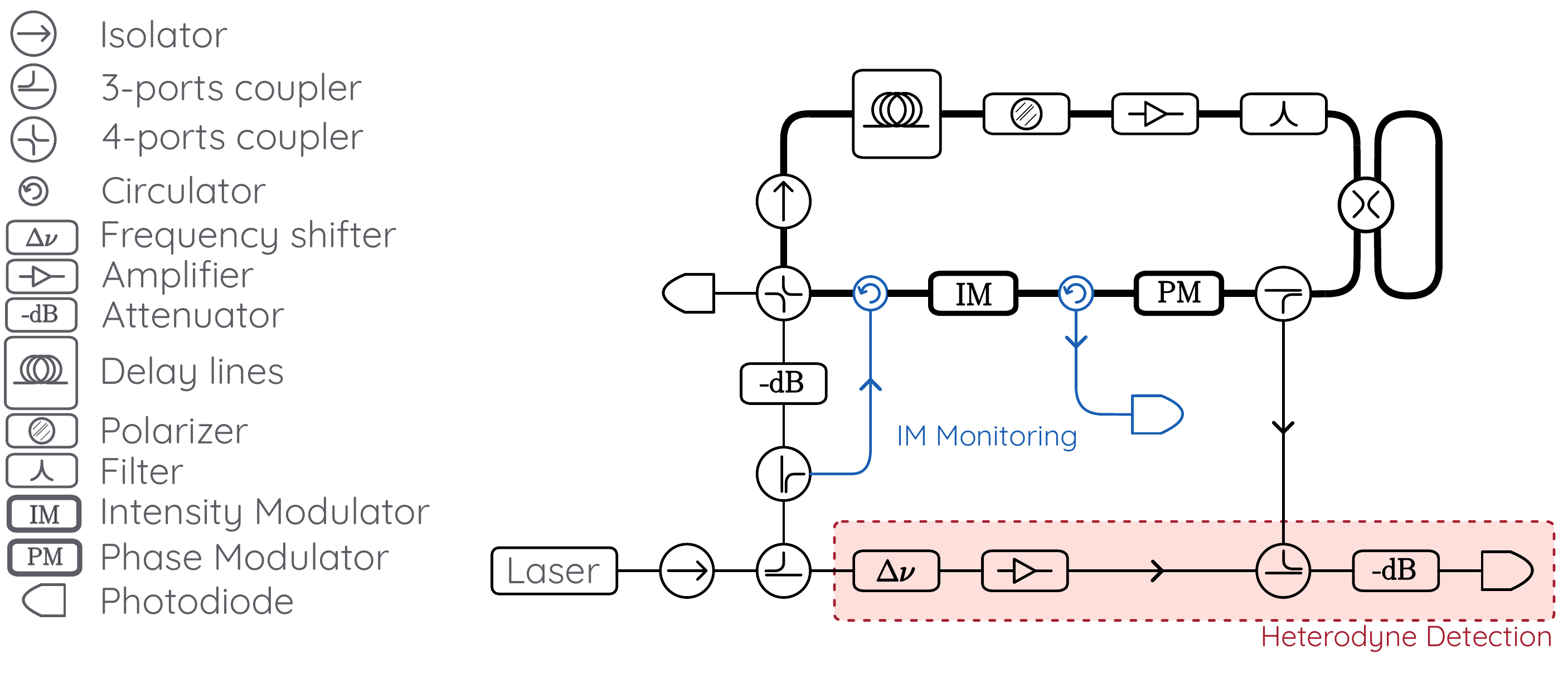}
\caption{Schematic representation of the experimental setup.}\label{fig:setupSchematic}

\end{figure}

A schematic of the the photonic platform used in this work is drawn in Fig.~\ref{fig:setupSchematic}. We use a NKT Koheras ADJUSTIK laser source with a central tunable wavelength at $1550.12~\unit{nm}$ and a linewidth $<100~\unit{Hz}$. An optical attenuator is used to limit the input drive amplitude and keep the setup in the linear regime. The output of the laser is split in three parts : a first part is guided to a heterodyne detection setup, a second is guided to the intensity modulator monitoring section, and the remaining power is injected in the main loop. A smaller secondary loop, around $0.3~\unit{m}$ long, is coupled to the main one via a 25:75 coupler. Two modulators are placed within the main optical fiber loop : a phase modulator (PM) and an intensity modulator (IM), each driven from a Zurich Instruments High Definition Arbitrary Waveform Generator (HDAWG). All optical components in the setup are polarization maintaining to ensure perfect alignment with the optical axes of the electro-optical modulators; a polarizer is used to further ensure proper polarization filtering. An Erbium Doped Fiber Amplifier (EDFA) is used to amplify the optical signal inside the main loop cavity in order to compensate for the losses caused by the insertion losses of the different components inside the loop and ensure a high quality factor ($Q\sim10^9$). A $100~\unit{GHz}$ bandwidth filter is used to suppress the amplification of undesired modes. This EDFA is further used as a source of noise (see below).
 
The internal Mach-Zehnder interferometer of the intensity modulator is prone to thermal fluctuations, leading to slow drifts (on the order of several minutes) of the operation point. To compensate these drifts, the state of the intensity modulator is monitored by injecting optical power in the opposite direction of the main propagation direction using two circulators around the modulator. A low bandwidth ($1~\unit{MHz}$) amplified photodiode signal monitors the transmitted optical power along this counter-propagating direction. This allows ensuring that intensity modulation is performed around mid-point, in a linear regime.
 
The heterodyne section consists of a $200~\unit{MHz}$ upward frequency shifter to provide a Local Oscillator (LO). This LO is then amplified with a Semiconductor Optical Amplifier (SOA) before being mixed with the probe signal via a coupler. The heterodyne signal is then attenuated before detection with a $600~\unit{MHz}$ bandwidth amplified photodiode. The photodiode signal is sampled by a $\SI{8}{\giga\hertz}$ bandwidth digitizer.

\subsection{Calibration protocol}

In order to compare different systems size with identical hopping strengths (and thus be able to obtain a relevant scaling), the hopping amplitudes $\alpha$ (phase modulation), $\beta$ (intensity modulation) and $\epsilon$ (perturbation, phase modulation) must be calibrated for each value of the main loop's FSR. Indeed, when changing the main loop's length, and thus its FSR, we change the frequency of the voltage signal sent to the EOMs to produce the $1^{st}$ and $3^{rd}$ nearest-neighbor couplings. Since the electro-optics coefficient linking the voltage to the index of refraction changes as a function of the driving frequency, the amplitude of each Fourier component must be adjusted in order to generate identical hopping strengths. To do this, we perform the following calibration procedure.

First, we open the auxiliary (small) loop such that no mode of the main loop hybridizes with it. Therefore we obtain an infinite lattice of modes separated by the main loop's FSR. We then apply only the nearest-neighbor phase modulation ($\alpha$), i.e. a Fourier component at exactly the FSR. This leads to a temporal band structure (as realized in previous implementations, including in Ref.~\onlinecite{dutt_experimental_2019}) in the form of a perfect cosine function. A small deviation of the driving frequency from the FSR results in an effective electric field and distorted bands (associated to Bloch oscillations). This allows perfectly calibrating the FSR and, by inspection of the resulting bands' amplitude, to estimate the strength of the reciprocal nearest-neighbor couplings. We then drive the phase modulator only with a Fourier component at 3 times the FSR ($3\Omega$). We still have a temporal band structure that is a perfect cosine function, but now with a period 3 times shorter (due to the increase of the lattice periodicity). The amplitude of this bands gives us a quantitative value of the perturbation $\epsilon$. 

Turning to the intensity modulation, we first calibrate its phase. This is important in order to account for any delay in the coaxial cables. Upon the presence of both the phase and intensity modulation, the band structure becomes asymmetric (see Fig.~2 (d) of the main text). We adjust the phase of the intensity modulation until the maximal intensity of the band structure occurs at the center of the rising edge to optimize the non-reciprocity. Importantly, we want the NHSE to accumulate toward lower frequencies. This allows accumulating the field, in frequency space, away from the LO frequency, and thus not having problems with negative frequencies in the heterodyne spectrum (see Fig.~\ref{fig:fftProcess} (c)). Finally, we close the auxiliary loop and now drive the cavity with the nearest-neighbor phase ($\alpha$) and amplitude ($\beta$) modulations to generate a NHSE in each N-site lattice. The amplitude of the intensity modulation is tuned until we obtain a specific slope of the field amplitude as a function of the lattice site (on a semilog vertical scale, see value chosen in the main text). This tuning of the NHSE, which is identical for all main loop's sizes, is chosen such that the total field amplitude circulating in the loop remains below a certain limit for the largest $N$, in order to prevent nonlinear dynamics. For the larger system sizes, this further requires lowering the input power in the loop. The responses reported in the main text are thus the output field normalized by the measured input field amplitude $|a_{in}|$.

\subsection{Heterodyne detection}

The main data of this work, reported in Figs.~3 and 4 of the main text, are taken from heterodyne spectra. This allows extracting the amplitude of the field in each frequency mode, notably at the edge of chains. As described above, heterodyne spectra are obtained by beating the signal radiating from the loop with a frequency shifted ($\SI{200}{\mega\hertz}$) copy of the laser. We accumulate the beating signal measured on a fast photodiode as the laser frequency is swept. In our implementation, contrary to the generic scheme depicted in Fig.~2 (b) of the main text, nonreciprocal transport was realized toward the lower frequencies to prevent the signals from running into the upconverted local oscillator. 

\begin{figure}
\centering
\includegraphics[width=0.9\linewidth]{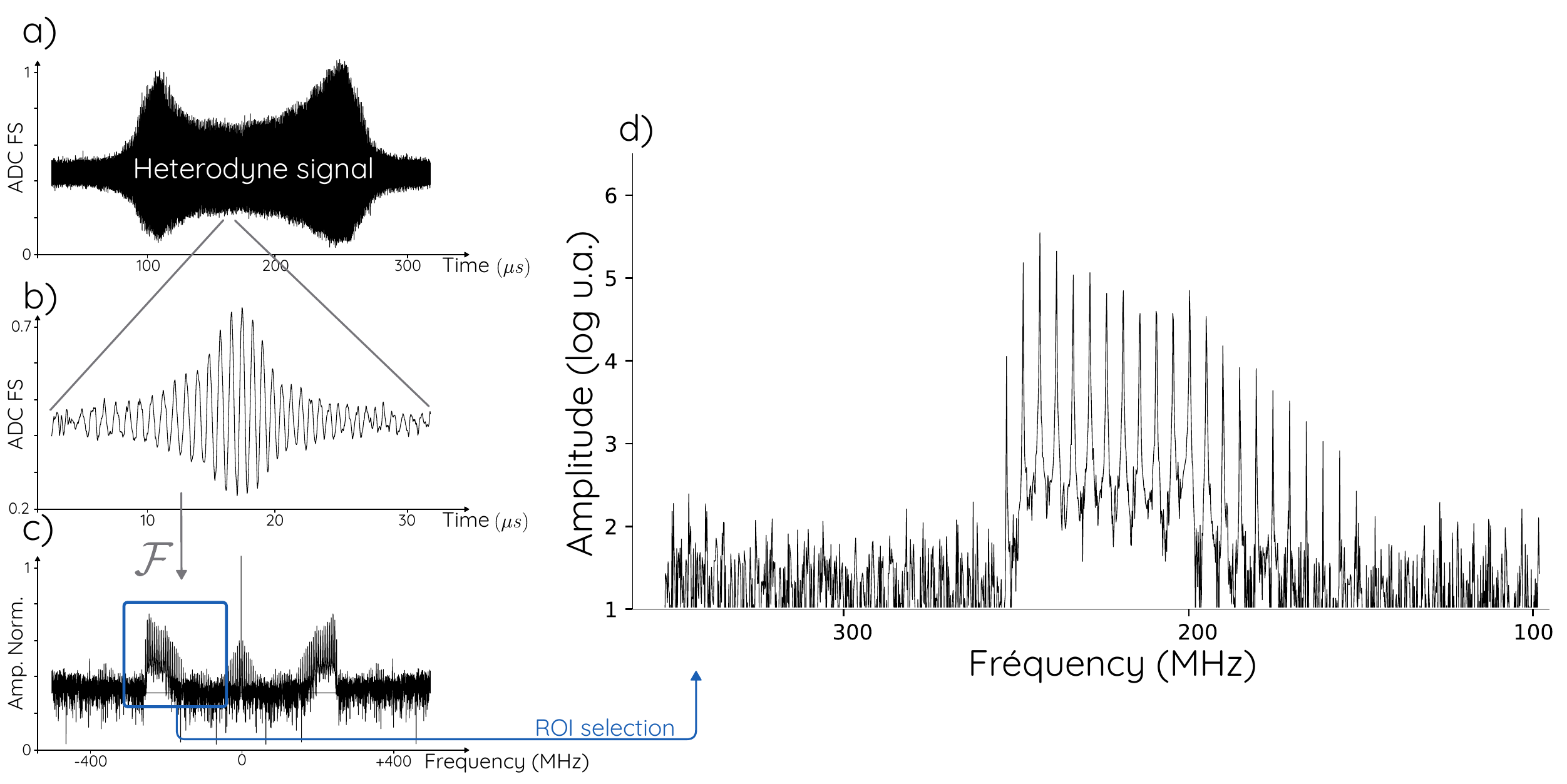}
\caption{(a) Time-resolved heterodyne signal as the laser detuning is scanned over a single band. (b) Sampling of the heterodyne signal around the region $\Delta=0$. (c) Fourier transfom of the sampled signal with a zoom-in (d) on a single chain exhibiting a clear NHSE in frequency space.}\label{fig:fftProcess}
\end{figure}

Figure~\ref{fig:fftProcess} (a) shows this heterodyne spectrum where we clearly see the profile of the band's density of states. From this spectrum, we can select a thin slice of $2^{18}\unit{pts}$ (b) around the center of the band to select data acquired with a minimal laser detuning. We finally perform a Fourier transform on this temporal region (c) with a resolution of $\sim 100~\unit{kHz}$ over $600~\unit{MHz}$. A zoom on a region of interest (ROI), for positive frequencies, shows a clear HNSE on a single chain (d). In this case, $\epsilon=0$ and the occupation of the neighboring chain is negligible.

\begin{figure}
\centering
\includegraphics[width=0.9\linewidth]{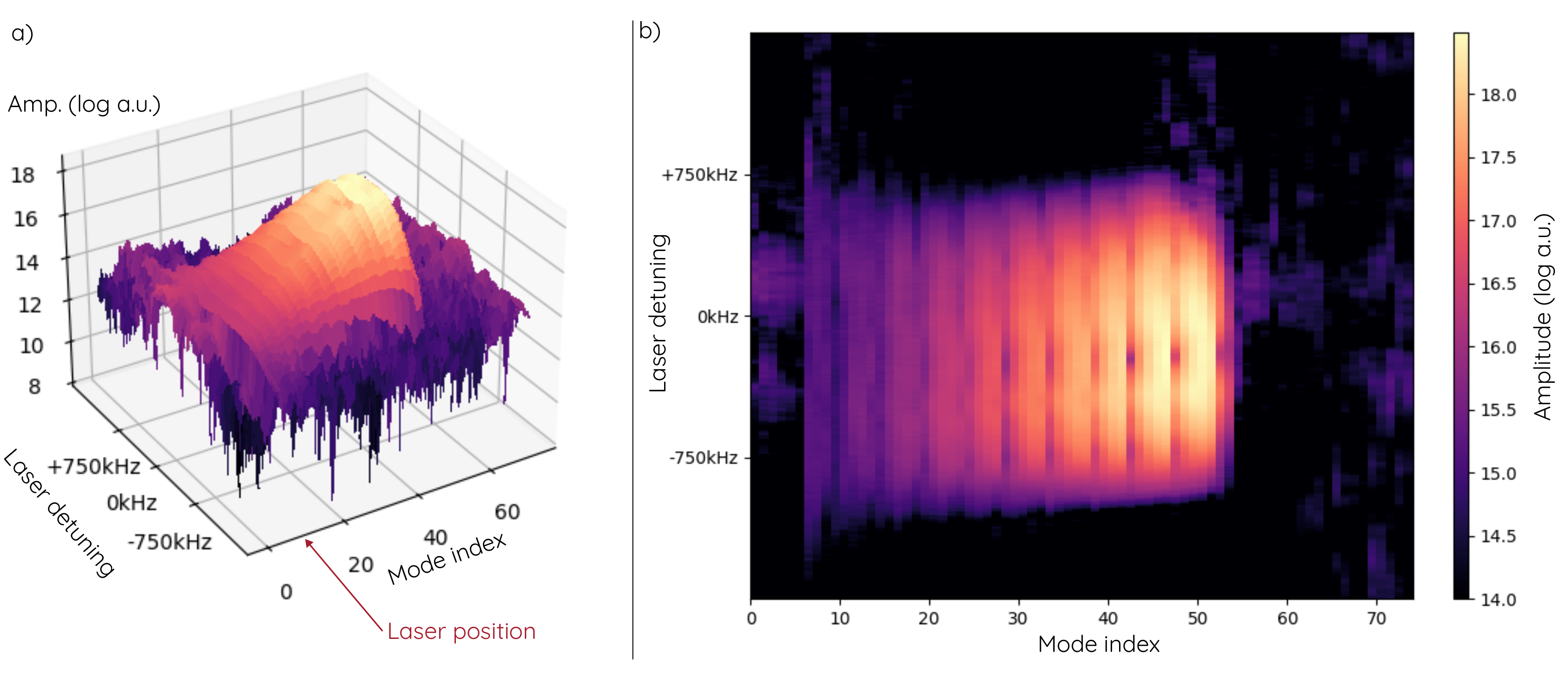}
\caption{Fourier transform of the heterodyne signal for slices of the temporal signal associated to distinct laser detunings over a single band. Panel (a) depicts a 3D mesh of this evolution, and panel (b) shows a top view. In both cases we see a clear NHSE in the pumped chain that evolves as a function of the laser detuning.}\label{fig:fftDetuning}
\end{figure}

In order to make sure that the sampled region in the heterodyne signal indeed corresponds to a region of minimal detuning, we can perform the same procedure for different slices of the full heterodyne signal. Figure~\ref{fig:fftDetuning} shows a heat map describing how the Fourier transform evolves as a function of the laser detuning, i.e. for different slices. We clearly see a maximal NHSE for $\Delta\rightarrow 0$, which allows identifying the ideal region for operating our system. This procedure is reiterated for each data presented in Figs.~3 and 4 of the main text to find the optimal operation point.

In the Supplementary Video appended, we show a summary of this procedure for every time window of the heterodyne spectrum measured for the case $N=34$. We clearly see that for cuts outside of the band structure, the Fourier transform only consists of noise contributions that populate every chain identically, because the driving laser is out of resonance and does not couple to the loop. Then, when the laser becomes resonant, we see a clear NHSE in both chains, the driven one and its neighbor, which become maximal when the laser detuning is minimal.

\subsection{Noise spectroscopy}

One key aspect of our work is the ability to probe the system's response to environmental noise. This is obviously critical for assessing the SNR. In order to extract this response, we remove the coherent drive. Hence, the only source of electromagnetic field entering the loop (beside the vacuum shot noise that is too faint to overcome our photodiode's electrical noise) comes from the spontaneous emission of the EDFA embedded in the cavity which operates below the lasing threshold (as described above). This emission fulfills the condition of a white, uncorrelated noise. We then measure the response of this input noise in the heterodyne signal (similarly as described above). This is how the noise data presented in Fig. 3 (b) and 4 of the main text was acquired.

\begin{figure}
\centering
\includegraphics[width=0.8\linewidth]{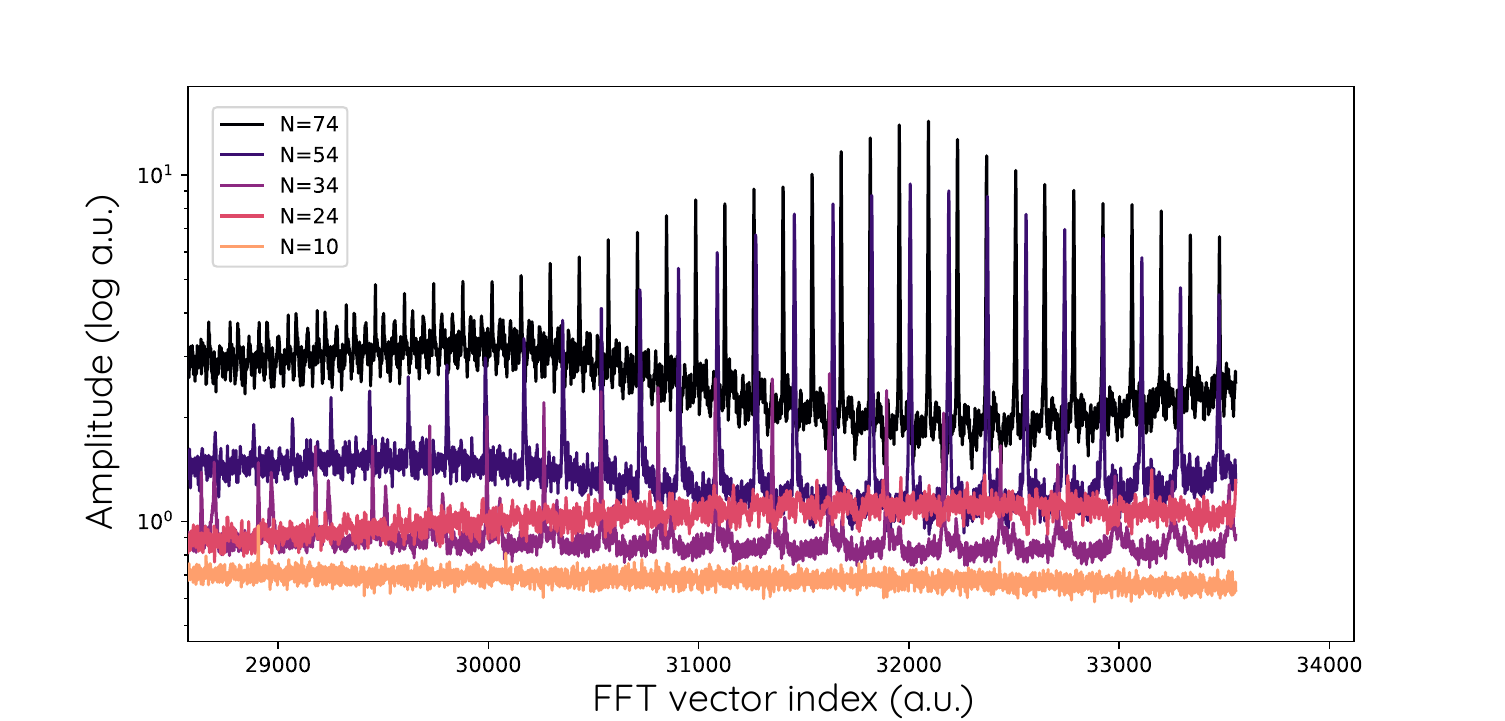}
\caption{Heterodyne spectra for different chain lengths where the input field is only incoherent noise. The occupation of the photonic modes is very clear, even though the laser is turned off, indicating that we indeed measure the photonic noise emitted by the EDFA.}\label{fig:noiseSpectroscopy}
\end{figure}

In order to distinguish this photonic noise from the electric noise of the detection system (photodiode + digitizer), we make sure that we see an occupation of the cavity modes of the loop. This indicates that the incoherent input field has indeed populated the modes of the cavity allowing to extract a scaling of this photonic noise specifically. Figure~\ref{fig:noiseSpectroscopy} shows a zoom-in of such heterodyne spectra where we clearly see the occupation of the cavity modes, except for the case $N=10$ that is dominated by the electrical noise (see Fig.~3 (c) of the main text).

\section{Numerical solutions of the Langevin equation}

In order to confirm the signal and noise scalings observed experimentally (e.g. in Figs.~3 and 4 of the main text), we have modeled our system using the following set of coupled Langevin equations :
\begin{equation}
    \dot{a}_{i} = i\left[H, a_{i}\right] - \frac{\gamma}{2}a_{i
    } - i\sqrt{\kappa}a_{in}^{(i)}
    \label{eq:langevin}
\end{equation}
where $a_{i}$ corresponds to the field amplitude in the $i^{th}$ frequency mode of the cavity, $\gamma$ is the optical lifetime, $\kappa$ is the input-output coupling strength and $a_{in}^{(i)}$ is the driving field amplitude localized on the $i^{th}$ site. The Hamiltonian $H$ for a system consisting a succession of $m$ $N$-site HN lattices is given by the following $mN\times mN$ matrix:

\begin{figure}[h]
    \centering
    \includegraphics[width=0.6\linewidth]{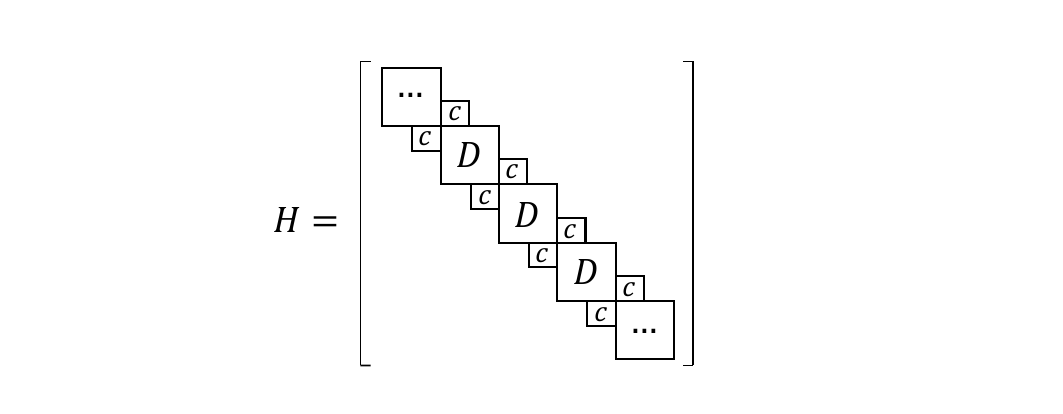}
    \label{fig:placeholder}
\end{figure}

\noindent where the $m$ $N\times N$ diagonal blocks 
\begin{equation}
D=\left[\begin{array}{ccccccccccc}
0 & J_{+} & 0 & \epsilon & 0  & \ldots & 0  & 0  & 0  & 0  & 0 \\
J_{-} & 0 & J_{+} & 0 & \epsilon & \ldots  & 0  & 0  & 0  & 0  & 0 \\
0 & J_{-} & 0 & J_{+} & 0 & \ldots & 0  & 0  & 0  & 0  & 0 \\
\epsilon & 0 & J_{-} & 0 & J_{+} & \ldots & 0 & 0  & 0  & 0  & 0 \\
0 & \epsilon & 0 & J_{-} & 0 & \ldots & 0 & 0 & 0  & 0  & 0 \\
\vdots & \vdots & \vdots & \vdots & \vdots & \ddots & \vdots & \vdots & \vdots & \vdots \\
0 & 0 & 0 & 0 & 0 & \ldots& 0 & J_{+} & 0 & \epsilon & 0\\
0 & 0 & 0 & 0 & 0 & \ldots & J_{-} & 0 & J_{+} & 0 & \epsilon\\
0 & 0 & 0 & 0 & 0 & \ldots & 0 & J_{-} & 0 & J_{+} & 0 \\
0 & 0 & 0 & 0 & 0 & \ldots & \epsilon & 0 & J_{-} & 0 & J_{+} \\
0 & 0 & 0 & 0 & 0 & \ldots & 0 & \epsilon & 0 & J_{-} & 0 \\
\end{array} \right]
\end{equation}
describe the dynamics of the particle inside a single HN chain and the $2\times 2$ off-diagonal blocks
\begin{equation}
C=\begin{bmatrix}
\epsilon & 0\\
0 & \epsilon
\end{bmatrix}
\end{equation}
describe the coupling between neighboring chains. Here, the $J_+$ and $J_-$ respectively describe the hopping amplitudes in the backward and forward directions, and $\epsilon$ the reciprocal $3^{rd}$ nearest-neighbor hopping amplitude inducing the perturbation coupling lattices. 

The use of a $mN\times mN$ Hamiltonian captures the fact that our system is invariant under translation in frequency space. The eigenspectra, as depicted in Fig.~2 (b) of the main text, consists of a concatenation of an infinite number of $N$-site lattices (in the approximation of negligible dispersion). In order to compare the simulations with our experimental data, we extract the amplitude of the field (the $a_i$) in two chains located in the middle of our simulated frequency space. This captures the essence of the experiment, where we only select a subset of two connected chains and ignore the ones at higher and lower frequencies.

To describe the dynamics of the system upon injection of white noise and of a coherent state (to model the noise and signal respectively), we propagate Eq.~\ref{eq:langevin} in time with the drive terms consisting either of a random, spatially uniform field (for noise) or a well-defined, localized coherent amplitude. 

In the first case, the random field is realized by generating a random array of driving fields $a_{in}^{(i)}$ (with $i$ ranging from $1$ to $mN$) by drawing, at each time step of the time evolution, random amplitudes and phases from a normal distribution such that, as explained in the main text,
\begin{equation}
    \langle a_{in}^{(i)}(t)a_{in}^{(j)}(t') \rangle = n_{th}\delta_{ij}\delta(t-t').
\end{equation}

In the second case, the coherent field is generated by using a driving field localized on the first lattice site of the first chain of the subset considered, defined in the rotating frame as:
\begin{equation}
    a_{in}^{(i)} = \delta_{i0}F_{i}e^{i\Delta_i t}.
\end{equation}
with $F_i$ the drive amplitude on the $i^{th}$ mode and $\Delta_i$ its detuning with respect to the $i^{th}$ mode. The factor $\delta_{i0}$ ensures that all the other lattice sites are undriven.

To confirm that the simulations are fully converged, we made sure that the results from the time evolution of Eq.~\ref{eq:langevin} with a coherent field matches its steady-state solution, i.e. solving the equations for $\dot{a}_{i}=0$. These converged simulations are those that were used for all theoretical calculations presented in the main text, including the SNR evolution as a function of $N$ (Fig.~3 (c)) and of $\epsilon$ (Fig.~4) where the calculated signal is presented in shaded blue and the calculated noise in shaded orange.

\begin{figure}[h]
    \centering
    \includegraphics[width=0.9\linewidth]{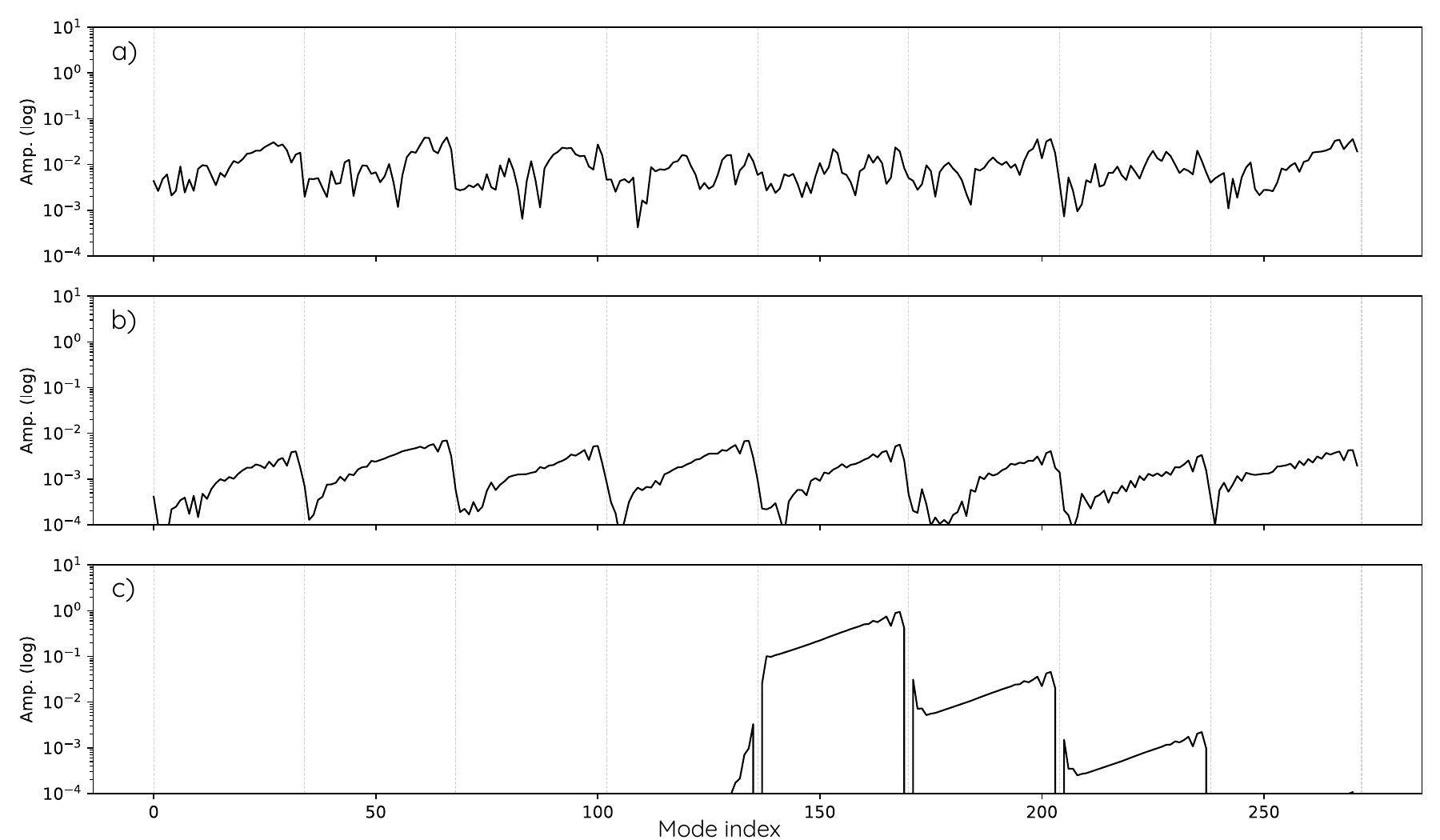}
    \caption{\textbf{Simulated steady state.} Each simulation was performed with the exact same conditions: $m=8$ chains of $N=34$ sites and $\epsilon=0.1 \frac{J_{+}+J_{-}}{2}$. a) Instantaneous state with only noise injected. b) Averaged steady state with noise. c) Coherent drive only, localized on the first site of the fifth chain. In each panel, the vertical dotted lines indicate the position of each chain.}
    \label{fig:simulated-profiles}
\end{figure}

The results of such simulations is presented in Fig.~\ref{fig:simulated-profiles} for $m=8$, $N=34$ and $\epsilon=0.1 \frac{J_{+}+J_{-}}{2}$. All simulations last $\SI{500}{\micro\second}$ and reach the steady-state within at most $\SI{250}{\micro\second}$. Panel (a) shows the instantaneous solution, i.e. for a given time increment, when the system is driven with white noise and for a time where the system has reached its steady-state. Panel (b) shows the same conditions, but a result that is time averaged over the last $\SI{250}{\micro\second}$ after the system has reached its steady-state, which allows a direct comparison with the result in the experiment (the heterodyne signal, as shown in Fig.~\ref{fig:noiseSpectroscopy}, is accumulated over a similar duration). In this latter case, one clearly sees the NHSE of noise which, on average, accumulates in every chain. The noise data points presented in Figs.~3 (b) and 4 (c) of the main text are obtained by extracting the maximal value in the chain where the signal is measured (see Panel (c)), and averaging this value over 5 simulation runs performed with five distinct noise realizations ; the width of the noise bands in Figs.~3 (b) and 4 (c) represent the standard deviation of these 5 simulation runs. The fluctuations  Panel (c) shows the solution of the simulation when the system is driven with a coherent drive localized on the first lattice site of the $5^{th}$ chain. We clearly see the accumulation in the pumped chain, as well as in its right-side neighbors; we further see the accumulation in a third chain that is reduced by a factor $\epsilon^{2}$.

\section{$e^{AN}$ scaling of the noise}

One of the main results of this work is the extraction of a scaling law for the noise in our system. This led to an exponential scaling of $e^{AN}$ where $N$ is the length of an individual HN lattice. As mentioned in the main text, this is distinct from the theoretical result proposed in Ref.~\cite{mcdonald_exponentially-enhanced_2020}. In that previous work, noise could only enter in the system through the first lattice site of a BKC (see Fig. 1 (b) of the main text). As a result, in the limit $\epsilon\rightarrow 0$, the only contribution to noise in the output field consists of thermal noise occupying the input-output waveguide because, in that limit of a vanishing perturbation, all noise entering the BKC can't change quadrature and can't thus be amplified through a round-trip along the chain. As a result, in that limit and under these conditions, noise does not scale with the system size; in a regime where $\epsilon$ is not vanishingly small, this result does not hold and the noise does depend on the lattice size $N$.

In our frequency-encoding scheme, as in Ref.~\cite{slim_optomechanical_2024}, noise can enter identically through all lattice sites. The above argument thus does not hold and we see a qualitatively different behavior. On key aspect that we observe is that noise follows exactly the same scaling as the amplitude profile of the NHSE in a single, isolated HN chain ($e^{AN}$). We can understand this scaling using the following argument.

Our system has translational invariance in frequency space, because we have an infinite succession (in the approximation of a non-dispersive medium) of HN lattices. This approximation is certainly valid over the relative low bandwidth considered in this work (typically $<\SI{1}{\giga\hertz}$). Over this bandwidth, we can well approximate the noise (spontaneous emission from the EDFA) as uniform, i.e. white noise. Hence, since we have an eigenspectrum and a noise source that are translationally invariant in frequency space, the noise profile in each chain must me identical by symmetry. This is confirmed in the experiment and in simulations (see Fig.~\ref{fig:simulated-profiles} (b)). Then, if we consider the processes by which noise can enter or exit a given chain that we label $i$, we have four contributions: A) noise entering/exiting from/toward the chain on the left, B) noise entering/exiting from/toward the chain on the right, C) noise emitted from the EDFA in each lattice site, and D) noise exiting the lattice through losses ($\gamma$) and the input-output coupler.

The rate at which noise enters or exits the $i^{th}$ chain through the first two processes is given by:
\begin{align}
    \Gamma_A &= \epsilon a_{N}^{(i-1)}-\epsilon a_{1}^{(i)}\\
    \Gamma_B &= \epsilon a_{1}^{(i+1)}-\epsilon a_{N}^{(i)}.
\end{align}

However, since $a_{N}^{(i-1)}=a_{N}^{(i)}$ and $a_{1}^{(i+1)}=a_{1}^{(i)}$ by translational invariance, those two contributions exactly cancel out. Therefore, the only net contributions come from processes C) and D) which are uniform in frequency and do not involve any neighboring chains. As a result the noise profile in each chain does not depend on the contributions from neighboring chains and scales identically as it would in a single, isolated chain, i.e. it follows the typical NHSE profile $e^{AN}$. This argument does not depend on the amplitude of $\epsilon$ and, as seen in Fig.~4 of the main text, the scaling of noise does not depend on that variable.

Finally, it is worth pointing out that if we had only one pair of HN chains, with noise uniformly entering each lattice site, as in the canonical model depicted in Fig.~1 (d) of the main text or in Ref.~\cite{slim_optomechanical_2024}, this argument would fail as we would not have a translationally invariant spectrum: noise entering a given chain from the left would not necessarily cancel out with noise exiting to the right. In the BKC picture, that would mean that noise entering in the P quadrature from the X quadrature is not canceled out by noise exiting the P quadrature to another degree of freedom. However, in the limit $\epsilon\rightarrow 0$, the contributions $\Gamma_{A,B}$ vanishes and, even in this configuration, we should retrieve a $e^{AN}$ noise scaling similar to the one we report in this work.

\section{Effect of $3^{rd}$ nearest-neighbor couplings in the bulk of the chains}

One key distinction between our implementation of the coupled HN chains and the canonical one presented in Fig.~1 (d) of the main text (and equivalent to the BKC proposed in Ref.~\onlinecite{mcdonald_exponentially-enhanced_2020}) is linked to the implementation of the perturbation linking the two chains. In the canonical model, only the last site of the first chain is linked to the first site of the second one. Here, as we explained in the main text, we added $3^{rd}$ nearest-neighbor couplings to couple the two chains, because the first and last lattice sites of each chain have low lifetimes. However, the perturbation we create does not only couple the last sites of the first chains to the first ones of the second chain. It couples, reciprocally, every mode of the chain to its $3^{rd}$ nearest-neighbor. It is important to mention that this modification does not qualitatively modify the behavior of the system. 

\begin{figure}
    \centering
    \includegraphics[width=\linewidth]{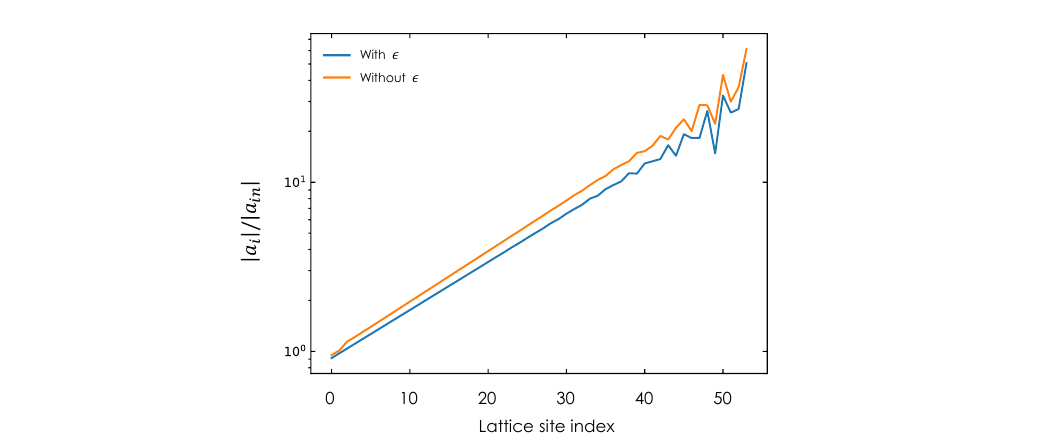}
    \caption{Steady-state solutions of the Langevin equation for chains without (blue) and with (orange) $3^{rd}$ nearest-neighbor couplings.}
    \label{fig:effect_of_epsilon}
\end{figure}

In Fig.~\ref{fig:effect_of_epsilon}, we show the steady-state solutions of Eq.~\ref{eq:langevin}, for a single chain, with and without $\epsilon$ (for $\epsilon=0.1\alpha$, the largest value used in this work). In the bulk of each chain, we observe a similar exponential localization, indicating that the NHSE is not significantly affected by the presence of this reciprocal perturbation $\epsilon$ coupling every $3^{rd}$ nearest-neighbors. We considered in these calculations a drive detuning $\Delta_{0}= 0.5\alpha$, similar to the one used for realizing Fig.~4 of the main text. Calculations showed that working in a perfectly resonant regime ($\Delta_0=0$, as was done for Fig.~3 of the main text) leads to a slight increase of the amplification factor $A$. The origin of this increased localization, for resonant excitations, is attributed to a modification of the trajectory $E(k)$ in the complex plane which slightly moves away from the origin and, in turn, decreases the skin depth. The impact of this perturbation on the bulk is an interesting point to explore in future works, as it could potentially lead to an enhanced SNR.

\section{Power spectra of the EDFA}

\begin{figure}
    \centering
    \includegraphics[width=0.9\linewidth]{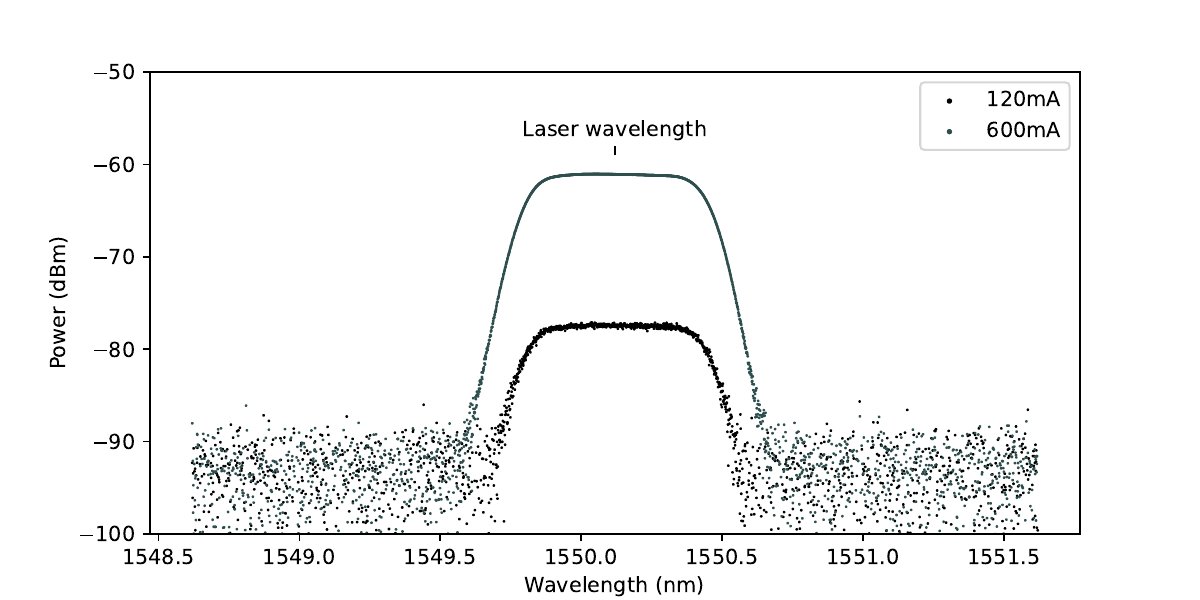}
    \caption{\textbf{Filtered EDFA power spectrum measured with an optical spectrum analyzer (OSA).} Measurement of the optical power spectrum of the EDFA filtered by the Fiberdyne Labs $\SI{100}{\giga\hertz}$ DWDM filter both used in the platform showing an homogeneous distribution. The power spectrum distribution is homogeneous at both low (high) pumping intensity, presented with black (gray) dots.}
    \label{fig:EDFA-OSA-spectrum}
\end{figure}

Figure~\ref{fig:EDFA-OSA-spectrum} shows the intensity spectrum of the ASE generated by the EDFA for two different levels of amplification. The spectra are highly homogeneous over a $\SI{100}{\giga\hertz}$ bandwidth and well centered on the laser frequency.

\end{document}